\documentclass[twocolumn,
]{ceurart}

\usepackage{url}

\begin{document}

\copyrightyear{2026}
\copyrightclause{Copyright for this paper by its authors.
  Use permitted under Creative Commons License Attribution 4.0
  International (CC BY 4.0).}

\conference{DCMI-2026 International Conference on Dublin Core and Metadata Applications}

\title{Hybrid Metadata Extraction from League of Nations Index Cards: From Feasibility Study to Archival System Integration}

\author[1,2,3]{Florian Cafiero}[%
orcid=0000-0002-1951-6942,
email=florian.cafiero@epita.fr,
]
\cormark[1]

\author[3]{Gr{\'e}goire Mallard}[%
orcid=0000-0003-4619-7251,
]

\address[1]{LRE, EPITA, Le Kremlin-Bic\^etre, France}
\address[2]{Centre Jean-Mabillon, \'Ecole nationale des chartes -- PSL, Paris, France}
\address[3]{Center for Digital Humanities and Multilateralism, Geneva Graduate Institute, Geneva, Switzerland}

\cortext[1]{Corresponding author.}

\begin{abstract}
This project report presents a hybrid AI-assisted workflow for extracting and reintegrating archival metadata from League of Nations index cards. The project is situated in the broader context of the Total Digital Access to the League of Nations Archives project (LONTAD). Rather than attempting full OCR of the underlying archival collections, the workflow targets the index cards themselves as documentary access points to files, series, archival descriptions, and digital objects. The project evolved from a layout-aware pipeline combining YOLO, TrOCR, and local LLM post-correction to a hybrid architecture using a fine-tuned vision-language model for broad extraction while retaining specialized OCR for file and series identifiers. 
\end{abstract}


\maketitle

Diplomatic archives are not only historical collections: they are also resources for studying the evolution of multilateralism, institutional decision-making, and international negotiations. Making such sources computationally accessible can support new forms of historical research, transparency, and public accountability \citep{Cafiero2023Datafying,CafieroCointetMallard2025Accountability}. In this context, metadata is not a secondary layer added after digitization; it is a condition for discovery, linking, interpretation, and reuse.

The Total Digital Access to the League of Nations Archives project (LONTAD), launched by the United Nations Library Geneva \citep{Wells2019LONTAD} is a major example of this shift. LONTAD digitized the League of Nations Archives while also addressing physical preservation, digital preservation, online access, and new pathways for digital humanities and data analytics research \citep{Leskinen2026LoNSampo,Hyvonen2026LoNDemo}.

The present project focuses on the League of Nations index cards preserved by the United Nations Library \& Archives Geneva. The project was carried out in collaboration with the Institutional Memory Section of the United Nations Office at Geneva (UNOG), with the aim of improving access to these archival references through metadata extraction, linking, and reintegration into archival information systems. The cards provide compact references to files, series, persons, organizations, topics, dates, and administrative structures. Their value lies in their role as documentary access points: they help users navigate from an index entry to archival descriptions, files, and digital objects. Extracting metadata from these cards is therefore not a generic OCR task, but a targeted contribution to the explorability of the League of Nations archives.

\section{Workflow}

\begin{figure}
  \centering
  \includegraphics[width=\linewidth]{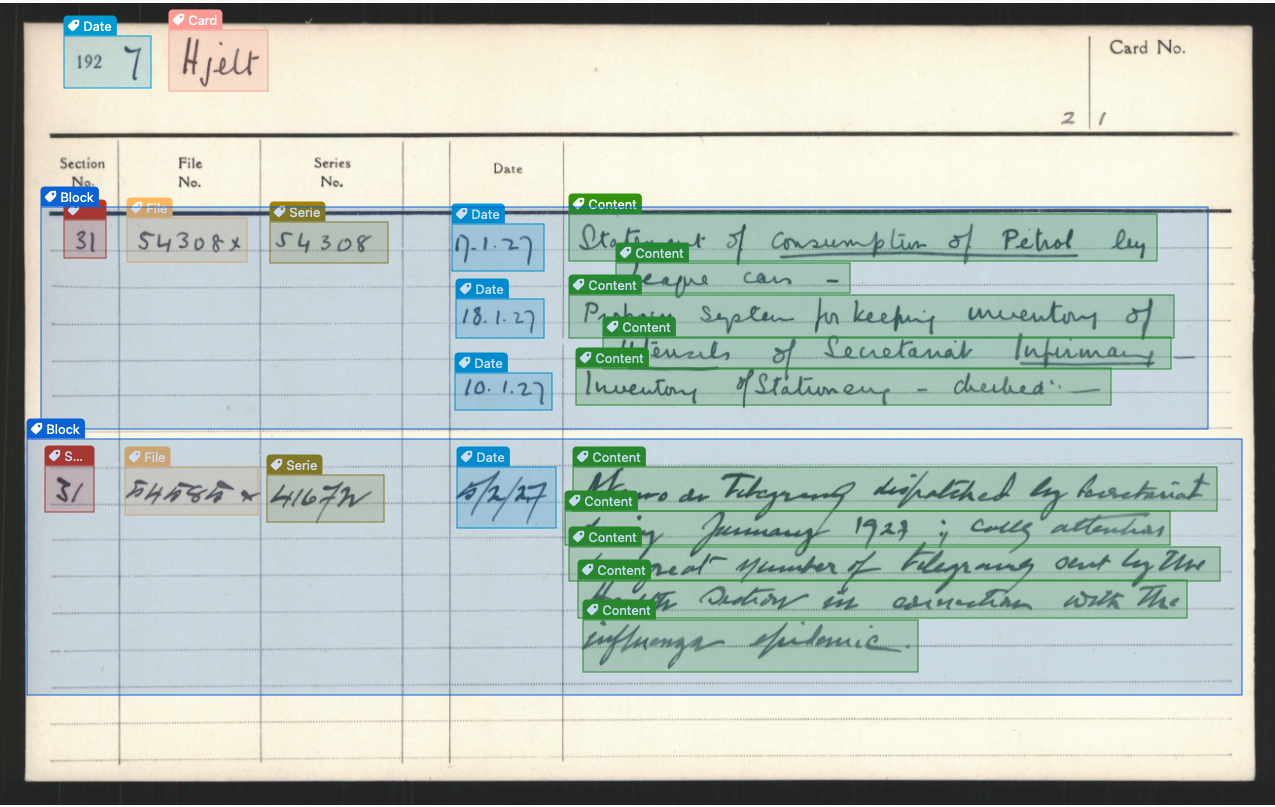}
  \caption{Example of the layout-aware detection stage on a League of Nations index card. Colored bounding boxes identify semantic fields such as date, card heading, block, section, file number, series number, and content. These detections supported block grouping, reading-order reconstruction, and field-level transcription in the first phase of the project. Source: League of Nations Archives, United Nations Library \& Archives Geneva; annotations produced within the project.}
  \label{fig:detection}
\end{figure}

The initial feasibility study, led with the Ecole nationale des chartes students, framed the task as targeted metadata extraction. A full OCR workflow over the underlying collections would have been more costly and uncertain because of heterogeneous layouts, varied material supports, multiple languages, and many personal and institutional names. The cards offered a more tractable target: they already condense archival references into a documentary interface used by readers and archivists.

The first implementation adopted a modular, layout-aware approach. A YOLO model was used to detect semantic fields on the cards \citep{Jocher2023YOLOv8}. Detected fields were grouped into logical blocks, and reading-order heuristics reconstructed the sequence in which the information should be interpreted. Text recognition was performed with a custom TrOCR architecture, combining a DeiT-base-patch16-384 visual encoder and an mBART-large-50 decoder \citep{Li2023TrOCR}. Approximately 500 randomly selected cards were fully transcribed by hand for fine-tuning. A locally deployed Mistral-family model was then used for post-correction \citep{Jiang2023Mistral7B}. This pipeline preserved field-level control and inspectability, but it was complex and exposed to error propagation across detection, grouping, reading order, transcription, and correction.

In a second phase, project requirements changed: precise field coordinates became less important than reliable metadata extraction and linking. This made it possible to move from field-by-field extraction to direct card-level extraction with a fine-tuned Qwen-family 3B vision-language model. The VLM-based workflow simplified the pipeline and improved general extraction quality. However, the improvement was not uniform across metadata fields. For broad descriptive extraction, the VLM approach was more effective; for short, high-value identifiers, especially file and series references, the earlier TrOCR-based component remained more reliable and less prone to hallucination. The resulting workflow is therefore hybrid: VLM extraction is used where semantic interpretation is useful, while specialized OCR is retained where exact recognition of constrained identifiers is essential.

\section{Metadata linking and system integration}

The project is now moving from extraction to archival system integration, with online availability expected by the end of June 2026. The remaining workflow includes compiling the CSV files delivered by the provider, extracting and organizing data by reference code, preparing CSV tables for import into AtoM, identifying with Artefactual the appropriate method for updating existing archival descriptions, updating these descriptions in AtoM, and ingesting the provider-delivered files into Preservica for long-term preservation.

The matching results provide operational indicators of the value of the workflow. According to the provider, \(87\%\) of the URLs present in the supplied database were matched with a reference found on an index card. In the opposite direction, \(40\%\) of the references present in the complete set of cards were matched with a URL. This second figure must be interpreted cautiously, because the proportion of card references for which a corresponding URL actually exists in the supplied database remains uncertain. It should therefore not be read as extraction accuracy, but as a combined indicator of extraction, normalization, database coverage, and linkability.

\section{Lessons learned}

The main lesson of the project is that archival metadata fields should not be treated as equivalent prediction targets. A subject heading, a date, a personal name, a file number, and a series identifier have different documentary functions and different error costs. Some fields support discovery and interpretation; others act as keys for database linking and retrieval. The latter require more conservative extraction strategies.

Rather than choosing between modular OCR pipelines and end-to-end vision-language models in general terms, archival institutions could align extraction strategies with the function and risk profile of each metadata field. Reliability is not a single global property of a model, but a field-specific requirement shaped by archival use, system integration, and long-term preservation.

\begin{acknowledgments}
The authors thank the United Nations Library \& Archives Geneva and, in particular, the Institutional Memory Section of the United Nations Office at Geneva (UNOG), represented in this collaboration by Blandine Blukacz-Louisfert and Hermine Diebolt, for access to the League of Nations index cards and for their guidance on the archival context, data provenance, and institutional requirements of the project. We also thank Fabrice Arlot and Pierre-Etienne Bourneuf for their support and coordination. We thank Calfa for their support for developing the technical pipeline as well as students from the Ecole nationale des chartes for their contribution to the feasibility report.
This work has received support under the Major Research Program of PSL Research University ``CultureLab'', launched by PSL Research University and implemented by ANR with the reference ANR-10-IDEX-0001. Errors remain our own.
\end{acknowledgments}

\bibliography{sample-ceur}

\end{document}